\RequirePackage{fix-cm}

\documentclass[smallextended]{svjour3}       
\smartqed  
\usepackage{graphicx}
\usepackage{mathptmx}      
\usepackage{array, eucal, mathtools, xfrac,color}
\usepackage{booktabs, array}
\usepackage{amsmath}%
\usepackage{amsfonts}%
\usepackage{amssymb}%
\usepackage{natbib}
\newcommand{\Int}{\int\limits}
\journalname{J. Math Biol.} 
\begin{document}

\title{Three-dimensional flow in Kupffer's Vesicle
    \thanks{TDMJ is supported by a Royal Commission for the Exhibition of 1851
        Fellowship. DIB was funded in part by a Bridgewater Fellowship and a
        London Mathematical Society Undergraduate Research Bursary to TDMJ. SSL
        is funded by an FCT Junior Investigator Fellowship and by an
    FCT-ANR/BEX-BID/0153/2012 grant.} 
}


\author{T. D. Montenegro-Johnson        \and
        D. I. Baker                     \and
        D. J. Smith                     \and
        S. S. Lopes
}


\institute{T. D. Montenegro-Johnson \at
              Department of Applied Mathematics and Theoretical Physics,
              University of Cambridge, UK \\
              \email{tdj23@cam.ac.uk}           
           \and
           D. I. Baker \at
              Department of Applied Mathematics and Theoretical Physics,
              University of Cambridge, UK \\
              \email{dib27@cam.ac.uk}
           \and
           D. J. Smith \at
               School of Mathematics, University of Birmingham, UK \\
               \email{d.j.smith.2@bham.ac.uk}
               \and
           S. S. Lopes \at
               CEDOC, Chronic Diseases Research Centre, NOVA Medical School \\
               Faculdade de Ci\^{e}ncias M\'{e}dicas, Universidade Nova de
               Lisboa, Campo M\'{a}rtires da P\'{a}tria, 130, 1169-056 Lisboa,
               Portugal \\
               \email{susana.lopes@nms.unl.pt}
}

\date{Accepted 13/01/2016, for J. Math Biol.}

\maketitle

\begin{abstract} 
    
Whilst many vertebrates appear {externally} left-right symmetric,
{the arrangement of internal organs is asymmetric}. In
zebrafish, the breaking of left-right symmetry is organised by Kupffer's Vesicle
(KV): an approximately spherical, fluid-filled structure that
{begins} to form in the embryo $10$ hours post fertilisation. A
crucial component of zebrafish symmetry breaking is the establishment of a
cilia-driven fluid flow within KV. However, it is still unclear (a) how dorsal,
ventral and equatorial cilia contribute to the global vortical flow, and (b) if
this flow breaks left-right symmetry through mechanical transduction or
morphogen transport. Fully answering these questions requires knowledge of the
three-dimensional flow patterns within KV, which have not been quantified in
previous work. In this study, we calculate and analyse the 
three-dimensional flow in KV.  {We consider} flow from
{both} individual and {groups} of cilia, and (a) find
anticlockwise flow can arise purely from excess of cilia on the dorsal roof over
the ventral floor, showing how this vortical flow is stabilised by dorsal tilt
of equatorial cilia, and (b) show that anterior clustering of dorsal cilia leads
to around $40\%$ faster flow in the anterior over the posterior corner. We argue
that these flow features are supportive of symmetry breaking through
mechano-sensory cilia, and suggest a novel experiment to test this hypothesis.
From our new understanding of the flow, we propose a {further}
experiment to reverse the flow within KV to potentially {induce}
situs inversus.  
     \end{abstract}

\section{Nodal cilia and symmetry-breaking flow}

In vertebrate embryos, the dorsal-ventral (back-front) and anterior-posterior
(head-toe) axes are the first to be established \citep{Hirokawa09}. The final
left-right axis then needs to be chosen in a consistent manner in order to allow
for asymmetric arrangement of internal organs. In humans for example, proper
establishment of the left-right axis leads to a heart on the left and a liver on
the right, despite apparent external bilateral symmetry. In certain species, the
establishment of left-right asymmetry is governed by a structure called the
node, first discovered in mice by \citet{Sulik94}. The fluid-filled node
expresses `nodal' cilia, which whirl in a clockwise direction when viewed from
tip to base. These cilia generate a fluid flow which plays a key role in
vertebrate left-right symmetry breaking \citep{Nonaka98}. 

Early theoretical and experimental studies of symmetry-breaking flow focused on
the mouse node. The embryonic mouse node is a triangular depression, covered
with a membrane and filled with fluid. The floor of the node is populated by the
whirling cilia which generate the internal fluid flow. Flow in the mouse node
was first modelled by \citet{Cartwright04}, who represented the motion of these
cilia by point torques driving an infinite fluid. When cilia were tilted in the
established posterior direction, clockwise whirling motion resulted in a
directional leftward flow, thereby breaking left-right symmetry. The predicted
tilt was subsequently observed experimentally by \citet{Okada05}. Other studies
used time-dependent cilium models \citep{Smith07,Smith08}, showing that particles
exhibit a leftward `loopy drift' when released just above cilia. Upon reaching
the left side of the node, particles recirculate slowly to the right just below
the upper membrane.

\begin{figure}[tb]
    \begin{center}
        \includegraphics{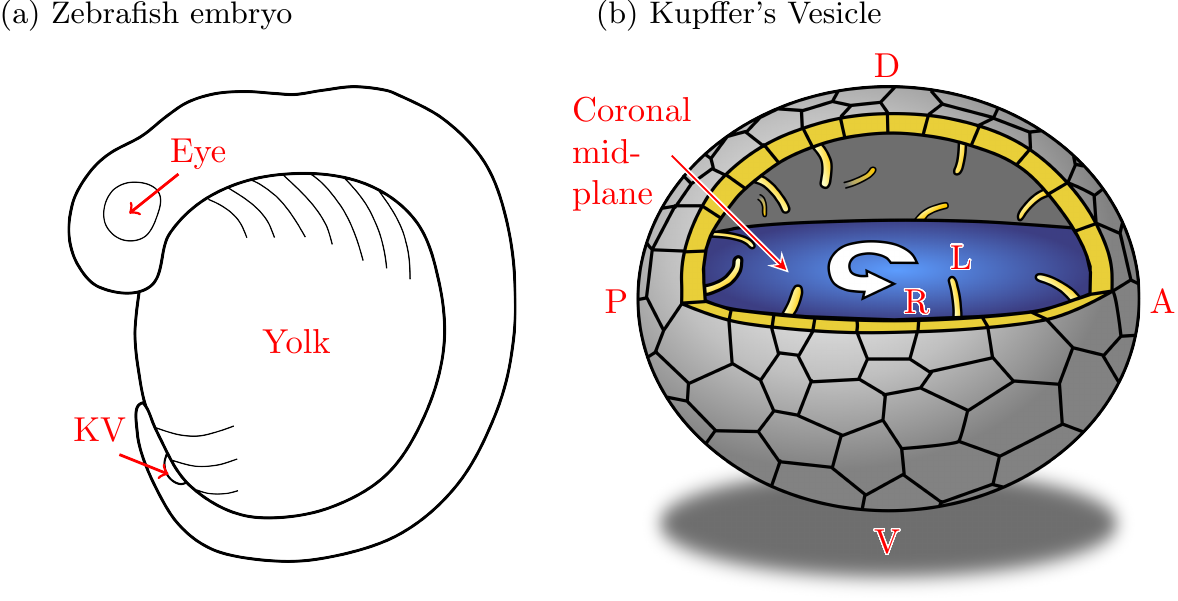}
        \caption{(a) Schematic of a zebrafish embryo, showing the approximate
            location of KV squeezed between the posterior and the yolk sac. (b)
            Cilia populate the inner surface of KV randomly, with a greater
            number on the dorsal roof and the highest density in the
            anterior-dorsal corner \citep[redrawn]{Kreiling07}. These
            cilia drive an anticlockwise vortical flow in the coronal midplane
            (midway between the dorsal and ventral poles).}   
            \label{fig:intro}
    \end{center}
\end{figure}

In the wake of experimental interest
\citep{Kawakami05,Kreiling07,Okabe08,Supatto08}, later studies began to examine
the organising structure in zebrafish (figure~\ref{fig:intro}a), known as
Kupffer's Vesicle (KV). KV is a transient structure that starts to form $10$
hours post fertilisation (h.p.f.), and when fully formed at $14$ h.p.f. (the $10$
somite stage) its architecture is more complex than that observed in many
species including mouse.  In live embryos it is approximately spherical, around
$50\,\mu\mathrm{m}$ across, and its entire inner surface is populated by cilia
which drive an internal flow (figure~\ref{fig:intro}b). These cilia are not
uniformly distributed; there are more cilia on the dorsal roof than the ventral
floor, with the distribution most dense (clustered) in the anterior-dorsal
corner \citep{Kreiling07}. In wildtype fish, around a fifth of these cilia are
immotile \citep{sampaio2014left}. The flow in the coronal midplane of KV is an
anticlockwise vortex when viewed from the dorsal roof, with a centre displaced
towards the anterior corner \citep{Supatto08}, though the fully
three-dimensional nature of the flow remains unknown.

\citet{smith2012symmetry} modelled flow in KV { using the
    regularised stokeslet boundary element method \citep{Cortez05,Smith09b},
    incorporating a time-dependent computational mesh of the full geometry and
    cilia. Flow in the coronal midplane was calculated at each timestep over $5$
cilium beats and then time averaged for comparison with experimental results
\citep{Supatto08}: a numerically intensive procedure taking approximately a day
of runtime.} It was found that dorsal tilt of equatorial cilia was required to
{best reproduce the experimentally-observed midplane} flow.
This model was extended by \citet{sampaio2014left} to account for natural
variation in cilium length, number, frequency and distribution in and between
embryos. The pattern of coronal flow was shown to be more robust in KV
with higher numbers of cilia.

However, it is still unknown how contributions from individual tilted cilia in
different locations on { the inner surface of KV} sum to produce a
3D flow field. As such, { the mechanism by which this flow breaks
left-right symmetry remains unclear}: (a) differential release/absorption of
morphogens or (b) mechano-sensory cilia. To address these questions a fully
three-dimensional description of the flow is required. In this work, we combine
the point torque modelling of \citet{Cartwright04} with a boundary element mesh
of the surface of KV, creating a hybrid singularity method. This approach bears
favourable quantitative comparison with time-resolved modelling and experimental
observations of flow speed and direction \citep{sampaio2014left}, and is able to
quickly evaluate the three-dimensional flow within KV.

\section{Modelling KV and nodal cilia}

\subsection{Fluid mechanics of Stokes flow}
\label{sec:fluid_model}

In KV, cilia are around $5\,\mu\mathrm{m}$ long, beating at around
$30\,\mathrm{Hz}$ in a roughly conical envelope with a semi-cone angle of around
$30$ degrees. Taking the highest velocity $U$ as approximately {the
length traced out by the cilium tip in a beat multiplied by the frequency,} $U \approx
2\pi\cdot5\sin{30}/(1/30) \approx 470\,\mu\mathrm{m}\cdot\mathrm{s}^{-1}$, we
see that the Reynolds number $\mathrm{Re}$ for flow in KV is 
\begin{equation} 
    \mathrm{Re} = 
    \frac{\rho U L}{\mu} \approx \frac{10^{3} \cdot 470\times 
    10^{-6}}{10^{-3}}\cdot 5\times10^{-6} \approx 0.0024 \ll 1, 
\end{equation} 
where $\mu,\rho$ are the dynamic viscosity and density of water respectively.
Since the Reynolds number is small, fluid flow driven by nodal cilia may be
modelled by the Stokes flow equations
\begin{equation} 
    \label{eq:stokes} 
    \mu\nabla^2\mathbf{u} -\nabla p + \mathbf{F} = 0, 
    \quad \nabla\cdot\mathbf{u}=0, 
\end{equation} 
for $\mathbf{u}$ the fluid velocity, $p$ the pressure and $\mathbf{F}$ body
forces acting on the flow {\citep{kim1991microhydrodynamics}}.

{We will model the time-averaged whirling of a nodal cilium by
a stationary point torque or ``rotlet'' 
$R_i(\hat{\mathbf{n}},\mathbf{x},\mathbf{y})$ in Stokes flow, which generates
the flow field $\mathbf{u}(\mathbf{x})$ \citep{Blake74a}
\begin{equation}
    u_i(\mathbf{x}) = M R_i(\hat{\mathbf{n}},\mathbf{x},\mathbf{y}) =
    M\frac{\epsilon_{ijk}\hat{n}_jr_k}{8\pi\mu r^3},\quad r_i = x_i - y_i,\quad
r^2 = r_1^2 + r_2^2 + r_3^2,
\end{equation}
where $\mathbf{y}$ is the location, $M$ is the strength, and
$\hat{\mathbf{n}}$ is the unit normal direction of the rotlet. Because the
Stokes flow equations~\eqref{eq:stokes} are linear, the flow from $n$ identical
cilia at locations $\boldsymbol{\chi}^n$ is then simply the sum of the
individual contributions
\begin{equation}
    \mathbf{u}(\mathbf{x}) = M\sum\limits_{n=1}^{N}
    \mathbf{R}(\hat{\mathbf{n}}^n,\mathbf{x},\boldsymbol{\chi}^n).
    \label{eq:rotlet_sum}
\end{equation}
In order to enforce the no-slip condition on the inner surface of KV, we also
require an integral \citep{Pozrikidis92} of the wall tractions $\mathbf{f}$ over
the boundary $D$
\begin{equation}
    \mathbf{u}(\mathbf{x}) = \underbrace{\Int_D \mathbf{S}(\mathbf{x},\mathbf{y}) 
    \cdot\mathbf{f}(\mathbf{y})\,\mathrm{d}S_y}_{\mbox{\scriptsize{no-slip
wall}}} + \underbrace{\vphantom{\Int_D}M\sum\limits_{n=1}^{N}
    \mathbf{R}(\hat{\mathbf{n}}^n,\mathbf{x},\boldsymbol{\chi}^n)}_{\mbox{\scriptsize{cilia}}},
    \label{eq:bem_with_rotlets}
\end{equation}
where $S_{ij}$ is the stokeslet tensor
\begin{equation}
    S_{ij}(\mathbf{x},\mathbf{y}) = \frac{1}{8\pi\mu}\left(
    \frac{\delta_{ij}}{r} + \frac{r_i r_j}{r^3} \right).
\end{equation}
The wall tractions $\mathbf{f}$ are unknowns, and are calculated through
specifying zero velocity on the surface of KV and solving the matrix system
arising from the discretisation of equation~\eqref{eq:bem_with_rotlets}. The
time-averaged velocity at any point $\mathbf{x}$ within KV is then found by
evaluating equation~\eqref{eq:bem_with_rotlets} with these tractions.  However,
for this model to represent flow inside KV,} the strength and direction of the
rotlets $M$ must be prescribed such that the flow field it generates matches the
time-averaged flow field of a {cilium} whirling about the axis
$\hat{\mathbf{n}}$. Thus, we now proceed with an analytical derivation for $M$,
and a validation of this model.

\subsection{Point torque cilium strength}

\begin{figure}[tb]
    \begin{center}
        \includegraphics{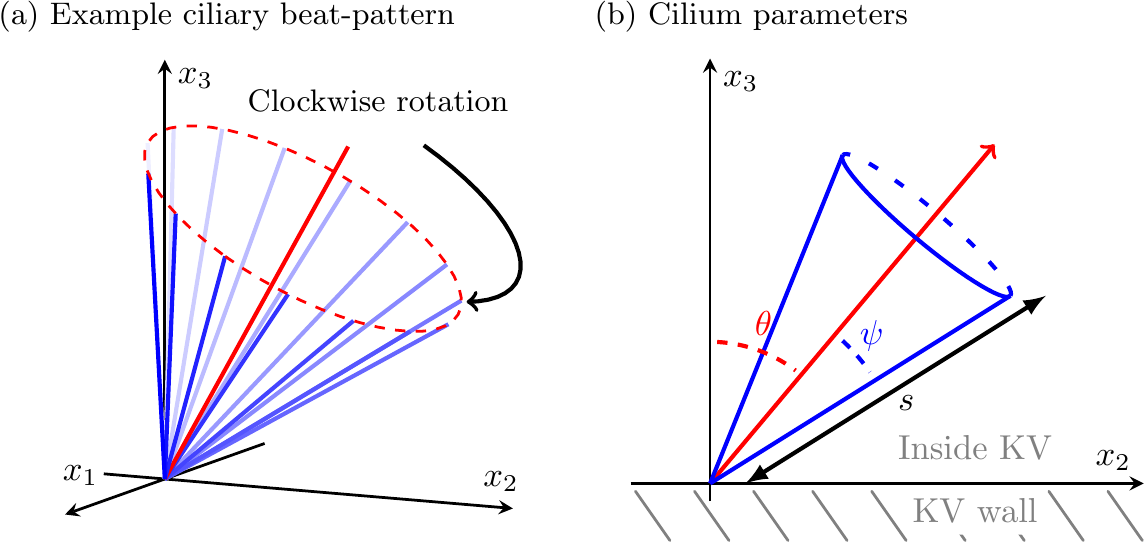}
    \end{center}
    \caption{Nodal cilia. (a) A ciliary beat-pattern described by equation
    \eqref{eq:beat_param} with semicone angle $\psi = 30^{\circ}$ and tilt angle
    $\theta = 30^{\circ}$, demonstrating clockwise rotation
    when the cilium is viewed from tip to base. (b) Schematic demonstrating the
    tilt and semicone angles $\theta,\psi$ respectively, showing the beat
    conical envelope and the measurement of the arclength $s$.}
    \label{fig:cilium_description}
\end{figure}

Following the methodology of \citet{Smith08}, we consider a straight,
tilted rod, tracing out a conical envelope (figure~\ref{fig:cilium_description}a), with
centreline $\boldsymbol{\xi}$,
\begin{subequations}
    \begin{align}
        \xi_1 &= \phantom{ - }s\sin\psi\cos\omega t, \\
        \xi_2 &= -s\sin\psi\sin\omega t\cos\theta - s\cos\psi\sin\theta, \\
        \xi_3 &= -s\sin\psi\sin\omega t \sin\theta + s\cos\psi\cos\theta.
    \end{align}
    \label{eq:beat_param}
\end{subequations}
The parameter $s$ is the arclength up the cilium, $\psi$ is the cilium
semicone angle and $\theta$ is the angle at which the cilium is tilted towards
the positive $x_2$ axis (figure~\ref{fig:cilium_description}b).

Using the local drag approximation of resistive force theory \citep{Gray55}, the
force $\mathbf{f}$ that the cilium exerts on the fluid is given by
\begin{equation}
    f_j = C_\parallel\left[\gamma\delta_{jk} - (\gamma -
        1)\frac{\partial\xi_j}{\partial s} \frac{\partial\xi_k}{\partial s}
        \right] \frac{\partial \xi_k}{\partial t},
\end{equation}
where
\begin{equation}
    C_\parallel = \frac{4\pi\mu}{2\log(2q/a)},\quad C_\perp = 
    \frac{8\pi\mu}{1 + 2\log(2q/a)},\quad \gamma = C_\perp/C_\parallel.
\end{equation}
The cilium diameter is given by $a$, and $q$ is such that $a < q < L$ for
cilium length $L$. The time-averaged moment per unit length that the cilium exerts on the
fluid is then,
\begin{equation}
    \langle \boldsymbol{\xi} \wedge \mathbf{f} \rangle = C_\perp\omega
    s^2\sin^2\psi \hat{\mathbf{n}}, \quad \hat{\mathbf{n}} =
    \cos\theta{\hat{\mathbf{x}}_3} + \sin\theta{\hat{\mathbf{x}}_2}.
\end{equation}
Integrating, the {magnitude of the} moment that the cilium exerts on the fluid is
\begin{equation}
    M = \Int_0^L {C_\perp\omega s^2\sin^2\psi\,\mathrm{d}s} =
    \frac{C_\perp\omega L^3\sin^2\psi}{3},
    \label{eq:equiv_rot_strength}
\end{equation}
{which gives the strength of the equivalent rotlet.}
Thus, we can model the cilium-induced time-average flow at any given point
$\mathbf{x}$ by a rotlet located at a point $\mathbf{y}$ of strength $M$
\begin{equation}
    \mathbf{u}(\mathbf{x}) = \frac{\mathbf{M}\wedge(\mathbf{x} 
    - \mathbf{y})}{8\pi\mu|\mathbf{x} - \mathbf{y}|^3}, \quad \mathbf{M} =
      {M}\hat{\mathbf{n}} = \frac{C_\perp\omega
      L^3\sin^2\psi}{3}\hat{\mathbf{n}}
    \label{eq:rotlet}
\end{equation}
{In appendix~\ref{sec:appendix}, the consistency of this
representation is demonstrated by showing that the equivalent rotlet over a
plane boundary generates the same volume flow rate per beat as the full time-dependent
model.}

Since the volume flow rate of the equivalent rotlet is independent of its height
above the boundary, we are free to choose its position in order to improve the
near-field approximation of the time-averaged flow arising from a beating
cilium. To achieve this, we consider a weighted average whereby the volume flow
rate produced by portions of the cilium below the rotlet is equal to that above.
{{The volume flow rate $Q$ from a point force ${f_1}$ in the
$x$-direction a
distance $d$ above a boundary is given by $Q\propto f_1 d$. In the
resistive force theory approximation, $|f| \propto |u| \propto s$, and since
$d\propto s$ then the volume flow rate per unit length is proportional to
$s^2$,}}
\begin{align}
    \Int_0^{d_r} \frac{f_1(s) \xi_3(s)}{\pi\mu}\,\mathrm{d}s &=
    \Int_{d_r}^{L\cos\psi} \frac{f_1(s)\xi_3(s)}{\pi\mu}\,\mathrm{d}s \nonumber \\
    \therefore\ \Int_0^{d_r} s^2\,\mathrm{d}s &= \Int_{d_r}^{L\cos\psi}
    s^2\,\mathrm{d}s \Rightarrow d = \frac{L\cos\psi}{\sqrt[3]{2}} 
    \approx 0.79L\cos\psi.
\end{align}
In appendix~\ref{sec:appendix}, we have compared the near-field flow of the
equivalent rotlet with the time-averaged flow of a time-dependent regularised
stokeslet model cilium. We find that in fact the best flow agreement is obtained
for $d = 0.82L\cos\psi$; for just half a length away from the cilium, the
relative error in the flow is less than $10\%$. This error decays quickly with
distance from the cilium, and the direction of the flow is consistent between
both models. Furthermore, since cilium-induced volume flow rate is proportional to $L^3$, a
$10\%$ difference in flow magnitude corresponds to a $3\%$ difference in the
length of the cilium. Such natural variation in cilium length occurs in KV
\citep{sampaio2014left}, accurate measurements of which are also subject to
limitations. Thus, to gain insight into the nature of the flow within KV, we
consider this level of accuracy acceptable.
The wall tractions $\mathbf{f}$ are unknowns, and are calculated through
specifying zero velocity on the surface of KV and solving the matrix system arising
from the discretisation of equation~\eqref{eq:bem_with_rotlets}. The
time-averaged velocity at any point $\mathbf{x}$ within KV is then found by
evaluating equation~\eqref{eq:bem_with_rotlets} with these tractions.

\subsection{Numerical implementation}

In order to model the structure of KV, we distribute rotlets within a boundary
element mesh of a sphere. The strength of these rotlets is set to be equivalent
to cilia of length $5\,\mu\mathrm{m}$ beating at $30\,\mathrm{Hz}$ with a
semicone angle of $30^{\circ}$. Rotlets are initially untilted, facing towards
the centre of the sphere, and the sign of the strength represents clockwise
rotation when viewed from the centre (ie, tip to base). Cilia are then tilted by
a specified amount in the local dorsal direction \citep{smith2012symmetry}.

Equation \eqref{eq:bem_with_rotlets} is discretised over a spherical mesh of
$512$ quadratic triangular elements; { the unknown traction is
modelled as taking the constant value $\mathbf{f}[l]$ over each mesh element
$E[l]$ of the surface $D$, so that the discrete version of
equation~\eqref{eq:bem_with_rotlets} is given by
\begin{equation}
    u_j(\mathbf{x}) = 0 =
    \sum\limits_{l=1}^{512}{f_i[l]}\int_{E[l]}S_{ij}(\mathbf{x},\mathbf{y})\,\mathrm{d}
    S_y + M\sum\limits_{n=1}^{N}
    {R}_j(\hat{\mathbf{n}}^n,\mathbf{x},\boldsymbol{\chi}^n),
    \label{eq:discrete_bem_with_rotlets}
\end{equation}
for $\mathbf{x} \in E[l]$. The element tractions $\mathbf{f}[l]$ are found by
solving the linear system~\eqref{eq:discrete_bem_with_rotlets}, and are such
that the velocity at the centroid of each element is zero.} Once the linear system is
solved, the element surface tractions can be used in the discrete form of
equation \eqref{eq:bem_with_rotlets} to calculate the fluid velocity at any
point within the mesh. Since time-averaging of the cilia beat has been
incorporated through the use of the equivalent rotlet, this matrix system need
only be solved once, taking a few seconds of runtime in contrast to the
time-dependent models where the matrix system is solved at each time-step. 

The code is implemented in Fortran 90 (gfortran, GNU Compiler Collection), with
mesh generation and boundary integrals performed using routines adapted from
BEMLIB \citep{Pozrikidis02}. The linear system is solved by LU factorisation
with the LAPACK routine \texttt{dgesv}. Flow visualisation is performed using
custom routines in Matlab, {with streamline data calculated using
the second-order variable two-step Adams Bashforth method
\citep{press2007numerical}} . Since to the authors' knowledge there is no known
general analytical solution for a rotlet inside a sphere, the method is verified
against the solution for a stokeslet in a sphere
\citep{oseen1927hydrodynamik,maul1994image} for which we find a relative error
{in the flow speed of $<1\%$ throughout the domain}. We now proceed
to analyse the flow in KV.


\newpage

\section{Results}

{ In the following results, the orientation of KV is consistent with
    figure~\ref{fig:intro}b. In figures with streamlines, the vortex direction
    is transparent to opaque and denoted by the large 3D arrows. Flow speed is
    indicated by streamline colour, and a lighting effect is used to show the 3D
    shape. The positions of rotlets, representing cilia, are given by the grey
    spheres. The lighter rotlets are closer to the right hemisphere. For each
    streamline figure, a corresponding video is available in the online
supplementary material.}

\subsection{Placement of ``useful'' cilia}

\begin{figure}[tb]
    \begin{center}
        \includegraphics{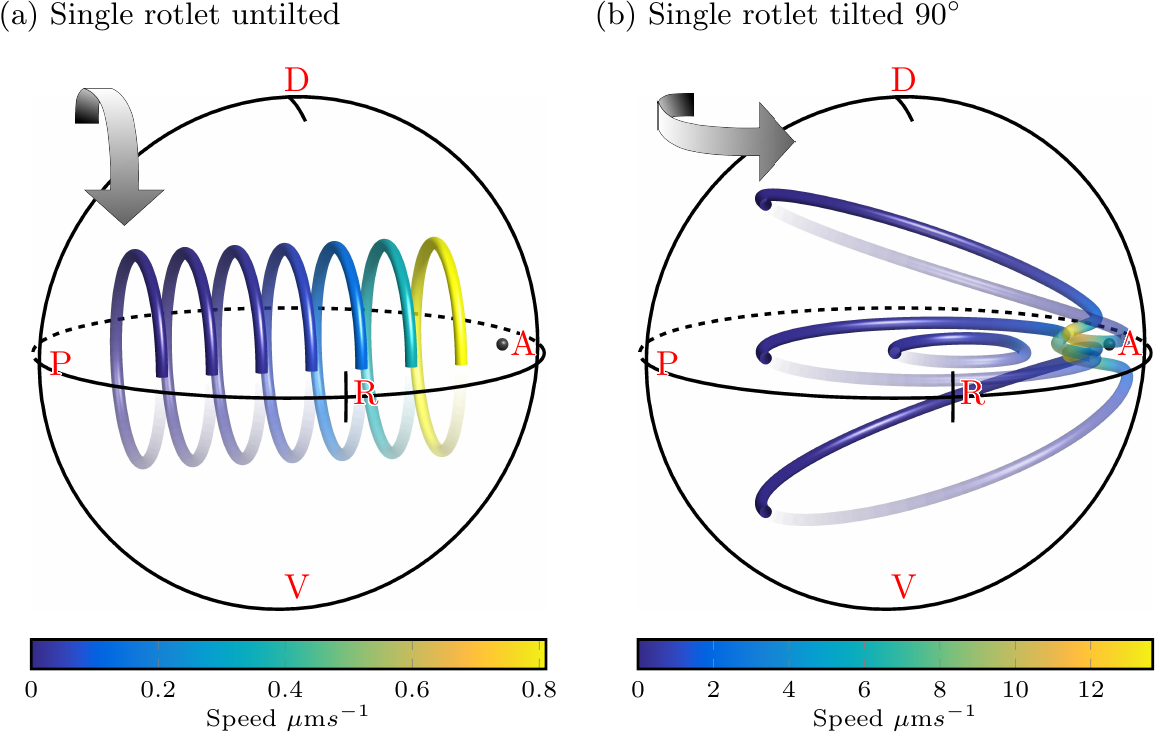}
    \end{center}
    \caption{Flows from a single {{rotlet}} located at the anterior
        equator, viewed from the right side of KV. (a) An untilted
        {{rotlet}}, showing vortical flow throughout KV.  (b) A
        $90^\circ$ tilt is applied to the rotlet, showing an anticlockwise
        vortex when viewed from the dorsal pole.  {By linearity of
        the Stokes flow equations, the} flow arising from equatorial cilia
    tilted by $30^\circ$ is a linear combination of these two flows.
{See the supplementary material for the videos corresponding to these
plots.}}    
    \label{fig:1cil}
\end{figure}

\begin{figure}[tb]
    \begin{center}
        \includegraphics{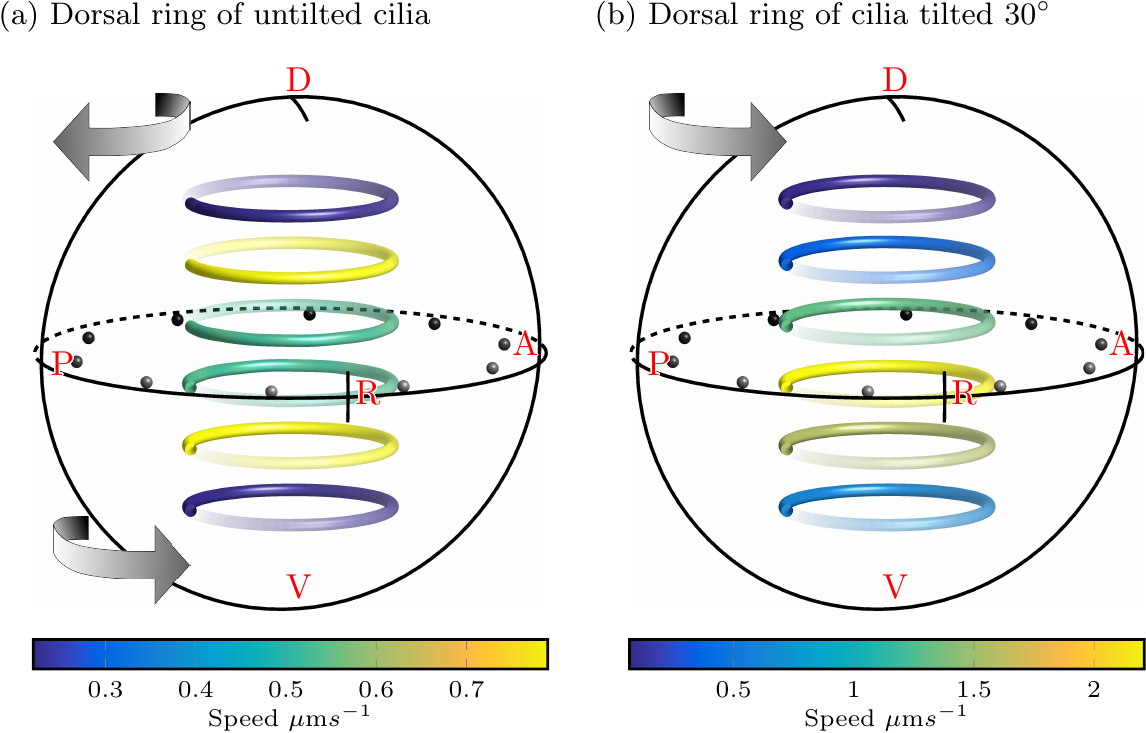}
    \end{center}
    \caption{Flows from a ring of equatorial cilia viewed from the right side of
        KV, with the same format as figure~\ref{fig:1cil}.  (a) Flow from an
        untilted ring of $10$ equatorial cilia, showing an anticlockwise vortex
        in the ventral hemisphere (bottom) and a clockwise vortex in the dorsal
        hemisphere (top). (b) After these equatorial cilia are tilted $30^\circ$
        to the dorsal pole, flow is an anticlockwise vortex throughout.
        {See the supplementary material for the videos corresponding to
            these plots.}}   
\label{fig:equatorial_ring}
\end{figure}

We begin our analysis by considering a single cilium placed at the anterior
equator, on the right side of each panel in figure~\ref{fig:1cil}. For an
untilted cilium, {the axis of the rotlet is perpendicular to the
wall and} flow is vortical (figure~\ref{fig:1cil}a); streamlines are concentric
rings about the anterior-posterior axis. Velocity is constant on each
streamline, denoted by the colour map, and decays as the distance away from the
cilium increases. The direction of the vortex is clockwise when viewed from the
posterior (left of figure), which as expected is the direction in which the
cilium rotates. This flow immediately shows us that untilted cilia located on
the dorsal roof (top of figure) are `useful' \citep{smith2014organized} for
generating the experimentally observed \citep{Supatto08} anticlockwise flow in
the coronal midplane (midway between the dorsal and ventral axes). Conversely,
cilia on the ventral floor are `antagonistic' to this flow, as they generate an
opposite whirlpool; a system comprising a single cilium at each of the dorsal
and ventral poles would have zero flow in the coronal midplane.
{Since the Stokes flow equations~\eqref{eq:stokes} are linear,
solutions may be superposed.} Thus, in the absence of cilium tilt, a surplus of
cilia on the dorsal roof is necessary to generate the observed anticlockwise
flow.

Untilted cilia on the dorsal roof contribute to the observed anticlockwise flow
in the coronal midplane, but equatorial cilia which are tilted towards the
dorsal pole also contribute. { Figure~\ref{fig:1cil}b shows flow
    arising from a rotlet aligned in the dorsal-ventral direction, parallel to
    the wall, which generates an anti-clockwise vortex. Because solutions to
    Stokes flow can be superposed, the flow from a tilted rotlet, representing a
    tilted cilium, can be thought of as a linear combination of the flow in
figure~\ref{fig:1cil}a and figure~\ref{fig:1cil}b}. This contribution motivates
consideration of a ring of equatorial cilia (figure~\ref{fig:equatorial_ring}).
Such a ring of untilted cilia induces (when viewed from the dorsal roof)
clockwise flow in the dorsal hemisphere, opposite to the naturally-occurring
flow, and anticlockwise flow in the ventral hemisphere
(figure~\ref{fig:equatorial_ring}a). Thus, equatorial cilia that are not tilted
in the dorsal direction are in fact antagonistic to anticlockwise vortical flow
in the ventral hemisphere.  However, once these cilia are tilted by around
$30^\circ$, similar to the average $26.6^\circ$ observed in mice
\citep{Nonaka05}, flow is anticlockwise flow throughout KV
(figure~\ref{fig:equatorial_ring}b), so that dorsally tilted equatorial cilia
strengthen the anticlockwise vortex.

\subsection{Natural cilium distribution and tilt}
\label{sec:anterior_cluster}

Before analysing the three-dimensional flow arising from a ``natural''
distribution of cilia, we draw comparisons with previous experimental data for
wildtype (WT) embryos \citep{sampaio2014left}. We consider the coronal midplane
flow generated by three random placements of $30$ cilia sampled from the
experimentally observed distribution of \citep{Kreiling07}, where $20\%$ of
cilia are found on the ventral floor, $17\%$ in the dorsal posterior corner,
$25\%$ in the dorsal mid-section and $38\%$ in the anterior-dorsal corner. Cilia
are tilted by an angle of $\theta = 30\sin(\alpha), \alpha \in [0,180]$ degrees
towards the dorsal pole, for $\alpha$ the cilium's latitude between the dorsal
and ventral poles. This way, equatorial cilia are tilted by $30^\circ$ and the
degree of tilt smoothly decreases to zero at either pole.

\begin{figure}[tb]
    \begin{center}
        \includegraphics{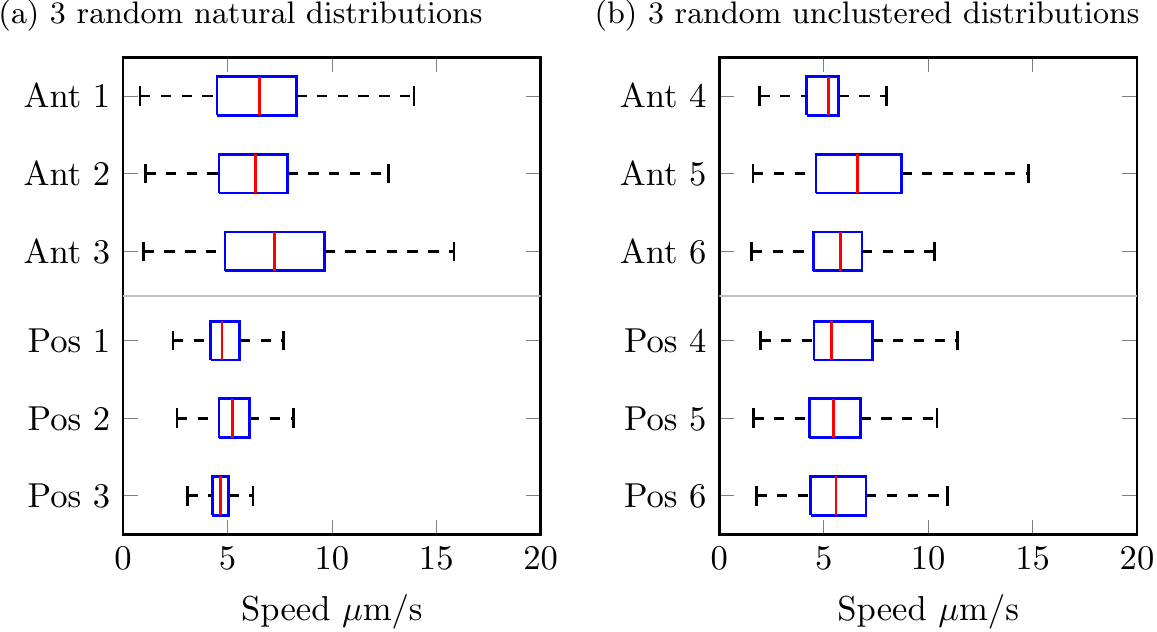}
    \end{center}
    \caption{Box plots of velocity magnitude sampled at $1500$
    random points in the anterior third and posterior third of the coronal
    midplane with $r \leq 17.5\,\mu\mathrm{m}$, i.e. at least half a length from
    any equatorial cilia. (a) The velocity for three separate random
    placements of cilia following the Kreiling distribution \citep{Kreiling07},
    showing consistently higher velocity in the anterior. (b) The velocity for
    three further random placements of cilia without anterior clustering, i.e.
    with $80\%$ of cilia on the dorsal roof and $20\%$ on the ventral floor,
    showing equal velocity in the anterior and posterior.}
    \label{fig:boxplots}
\end{figure}

Figure~\ref{fig:boxplots}a shows a box plot of the flow speed at $1500$ randomly
selected points in the anterior and posterior thirds of the coronal midplane for
three random natural placements of cilia. These boxplots show consistently
higher velocities, at around $40\%$, in the anterior when compared to the
posterior, and bear a striking resemblance to the data of Sampaio {et al.}
(Figure 2b). In particular the median velocity we find is $7\,\mu\mathrm{m}/s$
in the anterior and $5\,\mu\mathrm{m}/s$ in the posterior are similar to the
values of $9\,\mu\mathrm{m}/s$ in the anterior and $6\,\mu\mathrm{m}/s$ reported
from averaging $675$ particle tracks from $7$ embryos. A likely explanation for
the underestimation given by our code is that we have restricted ourselves to a
central region of the flow, at least $2.5$ lengths from any cilium where the
approximation of a time-averaged flow is valid, whereas particle tracks in
Sampaio {et al.} include near-cilium interactions where the flow velocity is
much higher. Furthermore, since Sampaio {et al.} follow native particles within
KV which originate at the surface, it is possible that a greater number of
particles are tracked nearer to the surface where flow is stronger, whereas our
sample is distributed evenly.

\begin{figure}[tb]
    \begin{center}
        \includegraphics{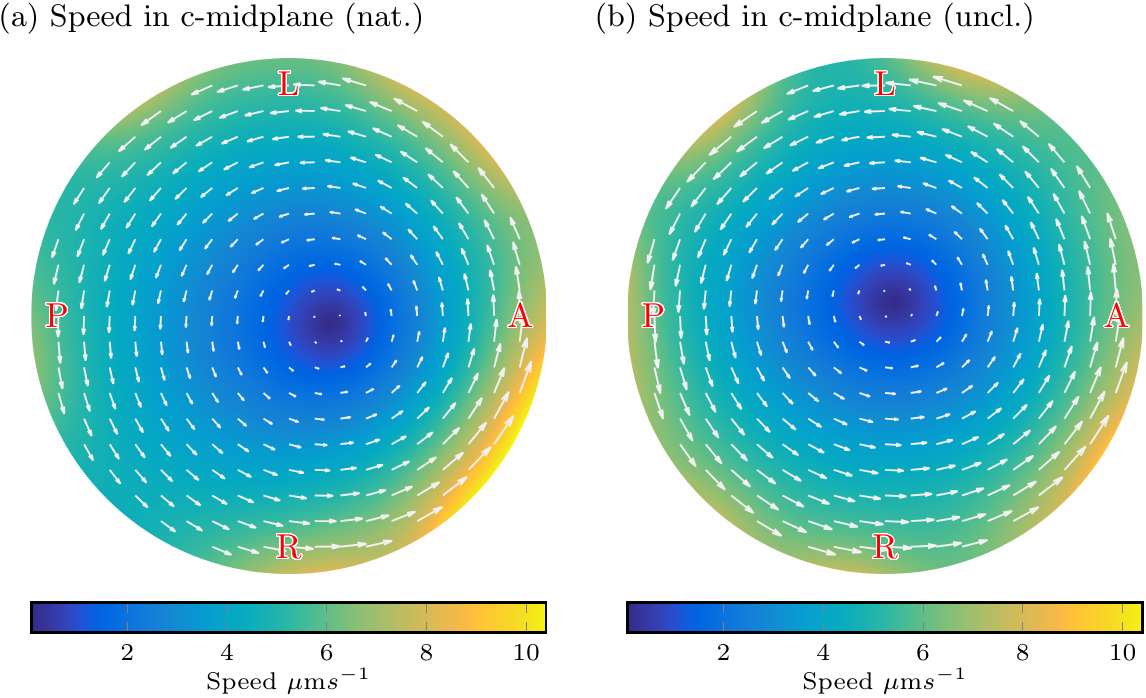}
    \end{center} 
    \caption{Flow speed in the coronal midplane for $r \leq 15$ for (a) a
        natural distribution (dist $1$), showing faster flow in the anterior
        (right) and a displaced vortex centre point, and (b) an unclustered
        distribution (dist 6), showing no significant anterior-posterior speed
        asymmetry and a vortex centre point located at the origin. A displaced
        centre point was observed experimentally by \citet{Supatto08}.} 
        \label{fig:c-midplanes}
\end{figure}

If the anterior clustering is disrupted, the difference between anterior and
posterior flow speeds disappears. Figure~\ref{fig:boxplots}b shows box plots of
the flow speeds in the anterior and posterior thirds of the coronal midplane for
three random cilium placements such that $20\%$ of cilia are on the ventral
floor and $80\%$ are on the dorsal roof; as observed in nature, but
without bias to the anterior-dorsal corner. Here we see no consistent difference
between anterior and posterior flows; the median flow speeds are
$5.5\,\mu\mathrm{m}/s$ in both the anterior and posterior. The effect of
anterior clustering upon the flow is also shown in figure~\ref{fig:c-midplanes}.
Figure~\ref{fig:c-midplanes}a shows midplane flow from a natural distribution
with anterior clustering. Flow is faster in the anterior, and the centre point
of the vortex is displaced to the anterior, as observed by \citet{Supatto08}. In
contrast, whilst there remains an anticlockwise vortical flow for the
unclustered distribution (figure~\ref{fig:c-midplanes}b), the vortex centre point
is no longer displaced. 

\begin{figure}[tb]
    \begin{center}
        \includegraphics{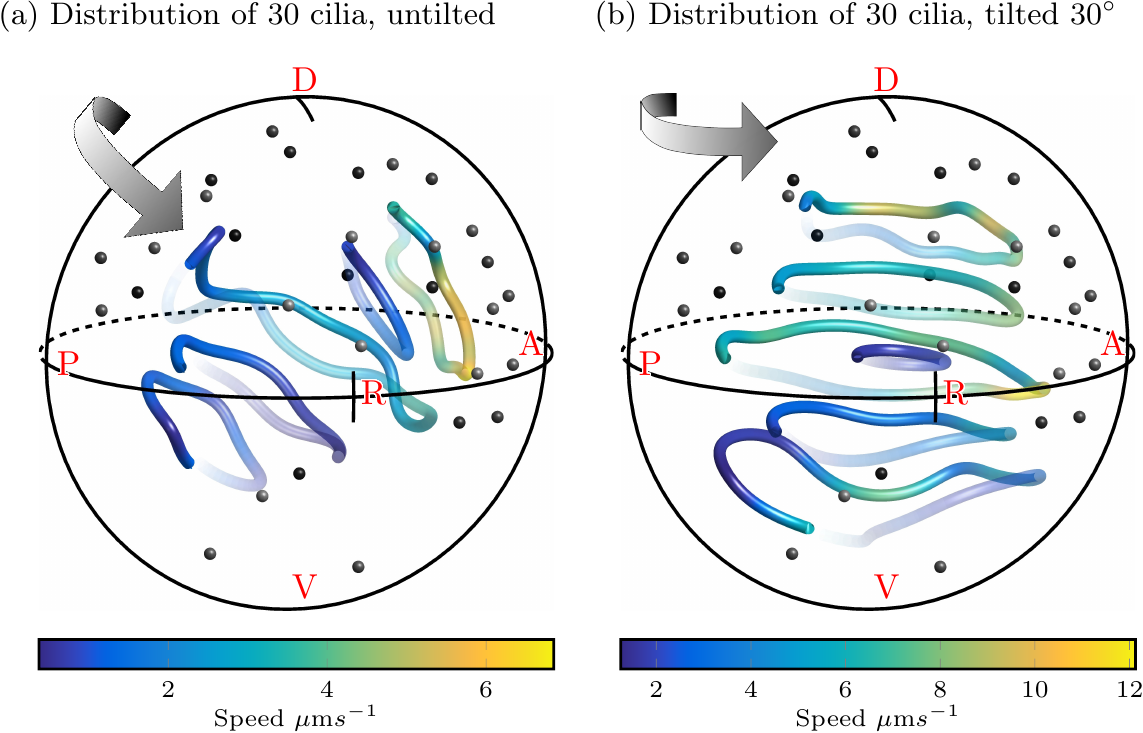}
    \end{center}
    \caption{Flows from a natural, random distribution of cilia viewed from the
        right side of KV. (a) Untilted cilia following the distribution of
        \citet{Kreiling07}, showing an anticlockwise vortical flow about the
        anterior dorsal corner and (b) the same placement of cilia tilted to the
    dorsal pole, flattening this vortex. {See the supplementary
material for the videos corresponding to these plots.}}
    \label{fig:natural_dist}
\end{figure}

We now proceed to examine the three-dimensional flow arising from a natural
distribution of $30$ cilia, {and the effects of dorsal tilt}. For
untilted cilia, flow is an anticlockwise vortex with a centreline pointing
towards the anterior-dorsal cluster, the location of the majority of cilia
(figure~\ref{fig:natural_dist}a). Such a flow would result in particles moving in
and out of the coronal midplane, which is not observed in experiments. However,
we then tilt the equatorial cilia towards the dorsal pole by an angle of $\theta
= 30\sin(\alpha), \alpha \in [0,180]$ degrees.  Figure~\ref{fig:natural_dist}b
shows that the effect of tilt is to flatten the vortical flow into coronal
planes. The centreline of the vortex is now aligned with the $z$-axis, and the
flow velocity is higher than in the untilted case because there is a greater
excess of `useful' cilia contributing to the final flow. Furthermore, flow
velocities are consistently higher throughout the anterior hemisphere than the
posterior hemisphere. These qualitative features were also consistent for
simulations with $20$ and $40$ cilia. 

{

\section{Proposed experiments}

Based on these results, and our new understanding of how cilia generate the 3D
flow-field in KV, we propose two novel experiments to (a) test the
mechano-sensory hypothesis by disrupting the clustering of cilia in the anterior
dorsal corner and (b) reverse the flow field to potentially induce situs
inversus.

\subsection{Anterior declustering}

The effect of anterior clustering of cilia is to increase the strength of the
vortex in the anterior corner, where the vortical flow is travelling leftward.
If a consistent vortex direction were sufficient to break symmetry, as might be
expected from morphogen transport/absorption, our results suggest that anterior
clustering would be unnecessary, provided a surplus of cilia were located on the
dorsal roof. Thus it seems plausible that both the direction of the vortex and
the relative strength between flow at the anterior and posterior edges of KV are
important. 

This observation motivates a novel experiment to examine the morphogen vs
mechano-sensory hypotheses as mechanisms for symmetry breaking. By selectively
knocking-out the motility of some cilia in the anterior cluster, perhaps through
laser ablation, it is possible to achieve an unclustered distribution, with a
dorsal surplus, of motile cilia within KV. These motile cilia would drive a
global anticlockwise vortical flow without a consistent speed difference between
the anterior and posterior hemispheres, as simulated in
figure~\ref{fig:c-midplanes}b. If such embryos were to exhibit significant situs
defects, this would support the presence of a mechano-sensory component to
symmetry-breaking, since for a pure morphogen transport mechanism the
anticlockwise direction of the vortex should be sufficient. However, it should
be noted that such a result would not rule out a morphogen-based component in
symmetry-breaking, as discussed further in section~\ref{sec:discussion}.}

\subsection{Flow reversal}

Since we now understand how some cilia can contribute to the vortical flow in KV,
while others are antagonistic, we can {furthermore} conceive a
non-invasive analogue of flow reversal experiments in mice \citep{Nonaka02} for
zebrafish: through knocking-out the motility of `useful' dorsal cilia leaving
only the antagonistic ventral cilia. In mice, embryos that developed following
flow reversal exhibited situs inversus, reversed positioning of internal organs.
Thus, such an experiment might provide valuable further insight into the
mechanism of symmetry breaking in zebrafish, particularly when taken together
with the experiment suggested in section~\ref{sec:anterior_cluster}. We now use
our model to examine the feasibility of generating a clockwise flow in KV.

\begin{figure}[tb]
    \begin{center}
        \includegraphics{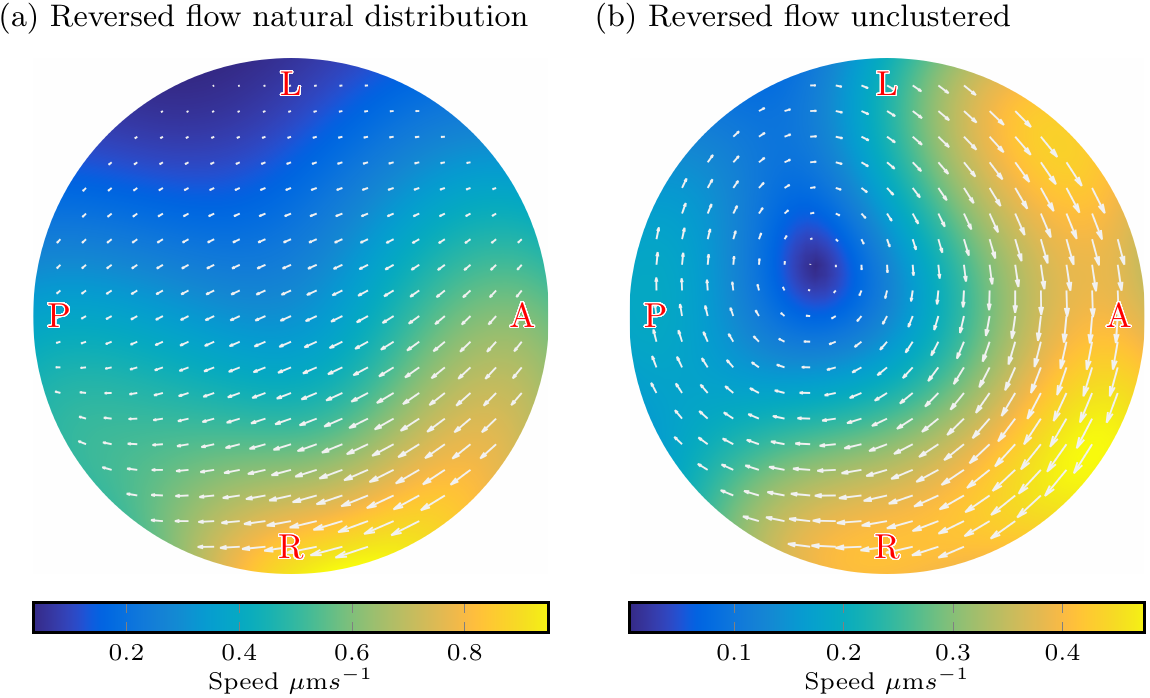}
    \end{center} 
    \caption{Coronal midplane flows from distributions of $40$ cilia for which
        only cilia in the bottom ventral third are motile. (a) Flow in natural
        distribution 2, showing slight directional motion and (b) flow in an
        unclustered, random distribution such as those created by
        \citet{wang2012regional} showing clockwise vortical flows
        in contrast with the naturally-occurring anticlockwise vortex
    (figure~\ref{fig:c-midplanes}a).}
    \label{fig:reversed_flow}
\end{figure}

Figure~\ref{fig:reversed_flow} shows two flow fields in the coronal midplane for
simulated embryos where only the cilia in the ventral quarter of KV are motile.
In figure~\ref{fig:reversed_flow}a, natural distribution 2 shows a more
directional than vortical flow in the majority of the midplane, which is clearly
a poor reversal of the anticlockwise vortical flow
(figure~\ref{fig:c-midplanes}a). Since there are very few cilia on the ventral
floor in the naturally-occurring cilium distribution, the location of these cilia
are not robust to random variations in precise placement. Thus, it is difficult
to ensure there are sufficient cilia in the correct location to ensure a global
clockwise vortex.

However, \citet{wang2012regional} showed that the process of differential cell
length to width growth which leads to cilium clustering in KV can be
disrupted by interfering with non-muscle myosin II activity. For an embryo where
cilia are evenly distributed throughout KV, a greater number of cilia are found
on the ventral floor, and the corresponding coronal midplane flow is a clockwise
vortex (figure~\ref{fig:reversed_flow}b). Thus, through either selecting wildtype
embryos with a large number of ventral cilia, or by disrupting cilium clustering
in the manner of \citet{wang2012regional}, it is possible to reverse the
vortical flow in KV by knocking-out the motility of all cilia except those on
the ventral floor. 

Note that for the particular random distribution chosen in
figure~\ref{fig:reversed_flow}b flow is stronger in the anterior than the
posterior. However, without clustering this effect is essentially random; whilst
the procedure above could reverse the direction of flow in KV, it is not
sufficient to ensure a reversal of the relative flow strength in the anterior
and posterior hemispheres. Faster reversed flow in the anterior might be
consistently possible, however, if a number of cilia in the ventral posterior
corner were ablated, or if equatorial cilia were somehow made to tilt towards the
ventral pole.

\section{Discussion}
\label{sec:discussion}

In this work, we have used a time-averaged singularity model that combines the
methods of \citet{Cartwright04} and \citet{Smith11} to examine
three-dimensional, symmetry-breaking flow in the zebrafish organising structure,
KV. An analytical formula was derived to determine the strength of a point
torque representing a cilium of given length, and at the optimum location this
point torque was shown to accurately represent the time-averaged flow generated
by a whirling cilium just half a length away. {While this model is
    valid for examining flow throughout the majority of the volume of KV, the time-dependent
    nature of the beat becomes important when analysing the motion of suspended
    particles closer to the cilium. Rather than a constant streaming, such
    particles have been shown in mice to execute a `loopy drift'
    \citep{Smith07,Smith08}. Furthermore, the time-averaged approach is not
    appropriate for studying dynamics in the chaotic layer between cilium tip
and KV wall \citep{Supatto08}. For such studies, the current model could be
modified to include time-dependent slender body representations of cilia using
the solution for a stokeslet within a sphere \citep{oseen1927hydrodynamik,maul1994image}.}
 
The flow from a single cilium was examined; { through linearity of
Stokes flow and superposition of solutions, this flow demonstrated} that a
surplus of cilia on the dorsal roof alone was sufficient to generate the
experimentally observed anticlockwise vortex. A ring of equatorial cilia was
antagonistic to this flow in the dorsal half of KV when untilted, but when
tilted in the dorsal direction added to the strength of the vortex. This
analysis suggests that flow reversal in KV might be achieved experimentally
through selectively ``knocking-out'' the motility of dorsal and equatorial
cilia, as achieved for left-sided cilia by \citet{sampaio2014left} .  This
hypothesis was tested, and was shown to be possible if the differential cell
shape mechanism responsible for cilium clustering was also disabled, as in
\citet{wang2012regional}. Simulated flow from natural distributions compared
favourably with previous experimental data, and dorsal tilt was shown in this
case to flatten the flow into a vortex about the dorsal-ventral axis.

What can this flow reveal about possible mechanisms of symmetry breaking in KV?
Two possible mechanisms for vertebrate symmetry breaking have been posited:
left-right differential concentration of morphogens responsible for breaking
symmetry, and mechanical sensing of the flow direction and strength by cilia. To
generate a non-uniform concentration of morphogen in KV, these must be
introduced and then reabsorbed at the surface through endocytosis. Morphogens
cannot be introduced with any inherent left-right asymmetry, and since the flow
is vortical in coronal planes throughout KV, morphogens introduced at the dorsal
and ventral poles would not be advected. Thus, any morphogen responsible for
symmetry breaking would have to be introduced at either the anterior or
posterior corners. Because flow is anticlockwise when viewed from the dorsal
roof, morphogen introduced at the anterior would first travel past the left side
of KV, whereas that introduced at the posterior would first pass the right.
Thus, if morphogens are reabsorbed sufficiently quickly (i.e. before reaching the
other side), a left-right differential concentration may be set up, thereby
breaking symmetry. 

However, it is not clear in this system why anterior clustering of cilia should
be necessary, as the mechanism should be equally valid for the unclustered
distributions in figures~\ref{fig:boxplots}b~and~\ref{fig:c-midplanes}b. If
higher velocity in the anterior corner relative to the posterior corner is
indeed required, it is supportive of at least a mechano-sensory component to
symmetry breaking; sensory cilia in the anterior would be deflected to the left
more strongly than cilia in the posterior were deflected to the right. We
proposed two experiments to control flow strength and direction in KV which
could help in the systematic analysis of zebrafish symmetry breaking. (1) By
selectively knocking-out the motility of a portion of cilia in the anterior
cluster of a wildtype embryo, an unbiased anticlockwise vortical flow can be
established. (2) By selectively knocking-out the motility of dorsal cilia in
embryos with homogeneous cilia distributions, an unbiased clockwise vortical
flow can be established. The results of such experiments would require careful
interpretation, particularly in the event of there being both mechanical and
morphogen-based elements to symmetry breaking, but could provide valuable additional
insight when applied in conjunction with other genetic tests on mutant and
knock-down embryos.

The numerical method presented is able to construct the time-averaged flow in KV
quickly and efficiently. From this model we have been able to gain further
understanding of the effects of individual cilia placement and tilt, and insight
into the way these effects combine to give the full flow, which was not possible
with two-dimensional visualisations. The nature of the three-dimensional flow
motivates a experimental means of flow retardation and reversal, which may
provide further evidence to the mechanism of symmetry breaking. {The
flexibility of describing the geometry with a boundary element mesh will further
allow for embryo-specific flow analyses in which the shape of KV and cilia
locations for any given embryo are extracted from imaging data.} The simple
analytical treatment of the cilium point torque strength suggests that this
method may in addition be useful in investigating other fish species with complex organising
structures,


\subsection*{Acknowledgements}

The authors would like to thank Julyan Cartwright and John Blake for continued
valuable discussion and insights, and Gabriele De Canio for supplying Matlab
code for a point force inside a sphere.

\begin{appendix}
    \section{Validation of rotlet cilium models.}
    \label{sec:appendix}

    {
We begin by validating the consistency of the equivalent rotlet model by
considering the volume flow rate induced by a tilted
rotlet located at $\mathbf{y}$ over a plane boundary \citep{Blake74a},
\begin{equation}
    u_i(\mathbf{x}) = \frac{M_j}{8\pi\mu} \Bigg[\frac{\epsilon_{ijk}r_k}{r^3} -
    \frac{\epsilon_{ijk} R_k}{R^3} + 2h\epsilon_{kj3} \left(
    \frac{\delta_{ik}}{R^3}-\frac{3R_iR_k}{R^5}\right) +
    6\epsilon_{kj3}\frac{ R_i R_k R_3}{R^5}\Bigg], 
    \label{eq:blake_rot}
\end{equation}
for $\mathbf{r} = \mathbf{x}-\mathbf{y}$, $\mathbf{R} =
\mathbf{x}-\mathbf{y}^\perp$ and $R^2 = |\mathbf{x} - \mathbf{y}^\perp|^2$ with
$h=y_3$ the distance between the singularity and the plane and
$\mathbf{y}^\perp=(y_1,y_2,-y_3)$. The volume flow rate $Q$ in the
$\hat{{x}}_1$-direction can be found by integrating the far-field terms,
\begin{align}
    u_i^{\mbox{\scriptsize{far}}} &= \frac{6\epsilon_{kj3}M_jx_ix_kx_3}
    {|\mathbf{x}|^5} + \frac{6h\epsilon_{ik3}M_3x_kx_3}{|\mathbf{x}|^5}, 
    \nonumber \\
    Q &= \Int_{0}^{\infty} \Int_{-\infty}^{\infty} u_1\,\mathrm{d}x_2
    \mathrm{d}x_3 = \frac{M\sin\theta}{2\pi\mu},
\end{align}
from which we see only the $\hat{{x}}_2$-component arising from the tilt,
$M\sin\theta$, contributes to volume flow rate in the $\hat{{x}}_1$-direction.
Substituting our equivalent rotlet strength \eqref{eq:equiv_rot_strength}, we
see
\begin{equation}
    Q=-\frac{M\sin\theta}{2\pi\mu} = \frac{C_\perp\omega
    L^3\sin^2\psi\sin\theta}{6\pi\mu},
\end{equation}
which matches the formula based on the force exerted by the whirling rod \citep{Smith08},
\begin{equation}
    Q = \frac{C_\perp\omega L^3}{6\pi\mu}\sin^2\psi\sin\theta.
    \label{eq:q_formula}
\end{equation}}

{In order to examine the accuracy of the equivalent rotlet model, we
now consider} a slender
    filament exhibiting kinematics given by equation \eqref{eq:beat_param}, and
    examine the flow using the method of regularised stokeslets. For a cilium
    above a plane boundary, the velocity at a point $\mathbf{x}$ in the fluid is
    given by 
    \begin{equation}
        \mathbf{u}(\mathbf{x}) = \int_S \mathbf{f}(\boldsymbol{\xi})\cdot
        \mathbf{B}^\epsilon(\mathbf{x},\boldsymbol{\xi})\,
        \mathrm{d}S_{\boldsymbol{\xi}},
        \label{eq:method_reg_sto}
    \end{equation}
    where $\boldsymbol{\xi}$ defines the cilium centreline and
    $\mathbf{f}(\boldsymbol{\xi})$ is the unknown force per unit length that the
    cilium exerts on the fluid. The tensor $\mathbf{B}^\epsilon$ is the
    regularised blakelet \citep{Ainley08,Smith11}, which incorporates image
    singularities with the regularised stokeslet to enforce no slip on the plane
    boundary
    \begin{align}
        \label{eq:regblakelet}
        B_{ij}^\epsilon(\mathbf{x},\boldsymbol{\xi}) =& \frac{1}{8\pi\mu}\bigg
        (\frac{\delta_{ij}(r^2 + 2\epsilon^2) + r_ir_j}{r_\epsilon^3} -
        \frac{\delta_{ij}(R^2 + 2\epsilon^2) + R_iR_j}{R_\epsilon^3} 
        \nonumber \\
        +& 2h\Delta_{jk}\left[ \frac{\partial}{\partial R_k} \left(
        \frac{hR_i}{R_\epsilon^3} - \frac{\delta_{i3} (R^2 + 2\epsilon^2) + R_i
        R_3}{R_\epsilon^3} \right) - 4\pi h \delta_{ik} \phi_\epsilon(R)\right]  
        \nonumber \\
        -& \frac{6h\epsilon^2}{R_\epsilon^5}(\delta_{i3}R_j -
        \delta_{ij}R_3)\bigg ),
    \end{align}
    where $\epsilon$ is a small regularisation parameter, chosen to be the
    radius of the cilium. We split the cilium into $20$ elements of constant
    force per unit length, and for each timestep collocate the known centreline
    velocity velocity at the centre of each element
    $\mathbf{u}(\boldsymbol{\xi}_i)$. This discretisation yields the linear
    system
    \begin{equation}
        \mathbf{u}(\boldsymbol{\xi}_m) = \sum_{n=1}^N\mathbf{f}_n\cdot
        \int_{S_n} \mathbf{B}^\epsilon(\boldsymbol{\xi}_m,\boldsymbol{\xi})\,
        \mathrm{d}S_{\boldsymbol{\xi}},
    \end{equation}
    which can be solved for the unknown force per unit length over each element.
    Having calculated the unknown force the cilium exerts upon the fluid at each
    timestep, we apply formula \eqref{eq:method_reg_sto} on a grid of points in
    the bulk flow, time average over a single beat, and compare them to the flow
    generated by a plane-boundary rotlet of `equivalent' strength
    \eqref{eq:blake_rot}.

    We begin by considering a cilium of length
    $5\,\mu\mathrm{m}$ and diameter $1/3\,\mu\mathrm{m}$, beating at
    $30\mathrm{Hz}$, which is typical of cilia found in KV. In the regularised
    blakelet model, this corresponds to a regularisation parameter of $\epsilon
    = 5/30$. The semicone angle $\psi = 30^{\circ}$, and the tilt angle $\theta
    = 30^{\circ}$. The cilium is tilted in the $y$-direction. We calculate the
    strength of the equivalent rotlet using both the analytical formula
    \eqref{eq:equiv_rot_strength} and the `volume flow rate matching' procedure detailed in
    appendix~\ref{sec:flux_match}. We find that using a normal force coefficient
    of the form \citep{Lighthill76}
    \begin{equation}
        C_\perp = \frac{8\pi\mu}{1 + 2\log(2q/a)}
    \end{equation}
    with $a = \epsilon$ and $q = L/3$ produces a relative difference between the
    equivalent rotlet strength as calculated by the two methods of less than
    $1\%$ for a wide range of cilium lengths and diameters, and thus we use the
    analytical form in these tests. 
    
\begin{figure}[tbp]	
        \begin{center}
            \includegraphics{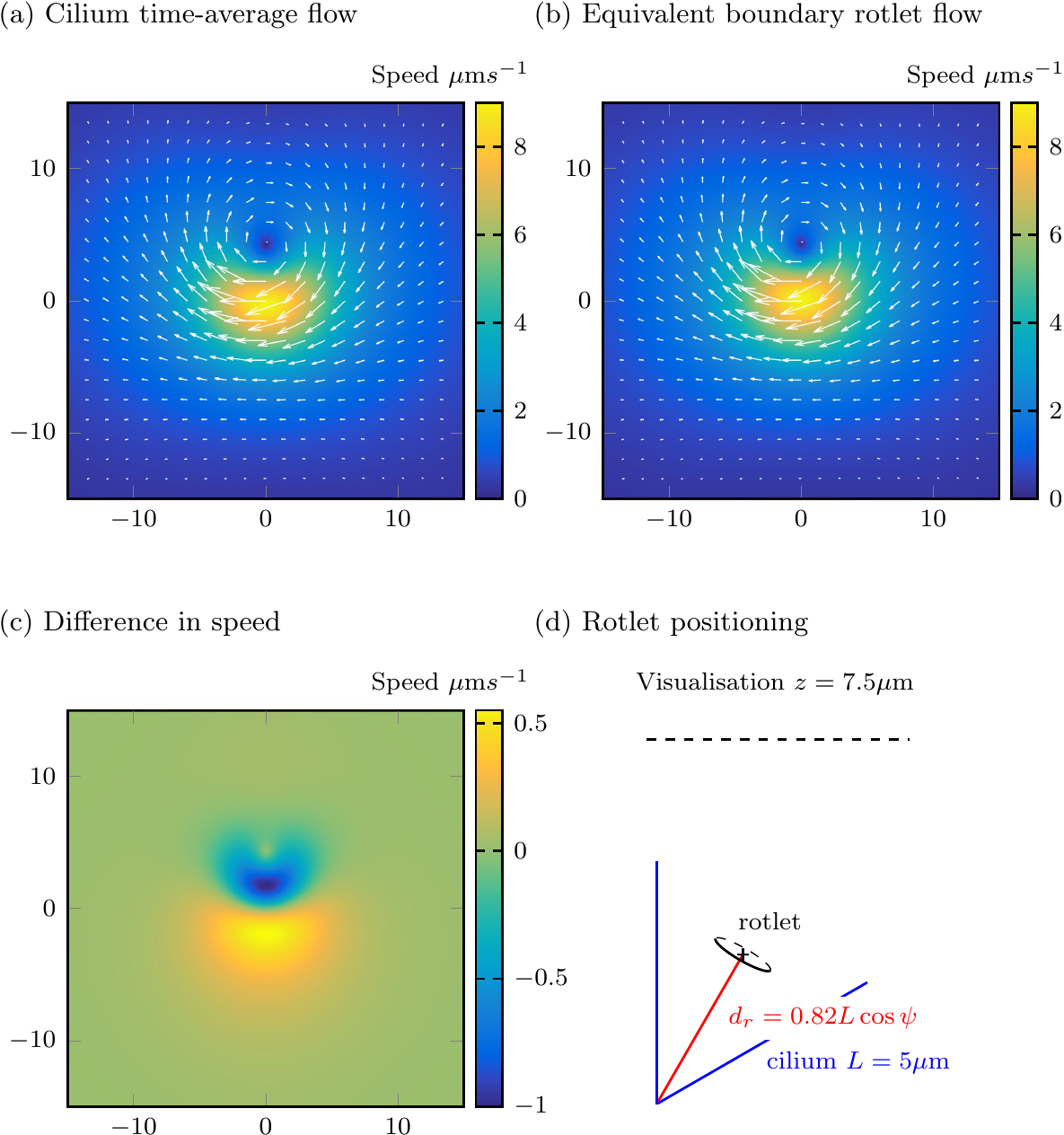}
            \caption{Comparing time averaged flow (a) of a $5\,\mu\mathrm{m}$ with
                $\psi = 30^{\circ}, \theta = 30^{\circ}$ beating at
                $30\mathrm{Hz}$ with the equivalent rotlet (b), in the plane $z
                = 7.5\,\mu\mathrm{m}$. Distances are shown in microns, and the
                colormaps in figures (a,b,c) are the speed of the flow in
                microns per second.  Visual comparison of (a) and (b) show
                remarkable similarity between the two solutions, with (c)
                showing at worst a $10\%$ error in the speed calculated by the
                rotlet model, and very good qualitative agreement in flow
                direction and magnitude. Evaluation of the results in (a) took
                approximately $2$ hours on a laptop computer, whereas (b) took
                approximately $10$ seconds. Panel (d) shows the envelope of the
                cilium, the location of the equivalent rotlet, and the location
                of the plane in which the velocity is evaluated.} 
            \label{fig:z_plane}
        \end{center}
    \end{figure}

\begin{figure}[tbp]	
        \begin{center}
            \includegraphics{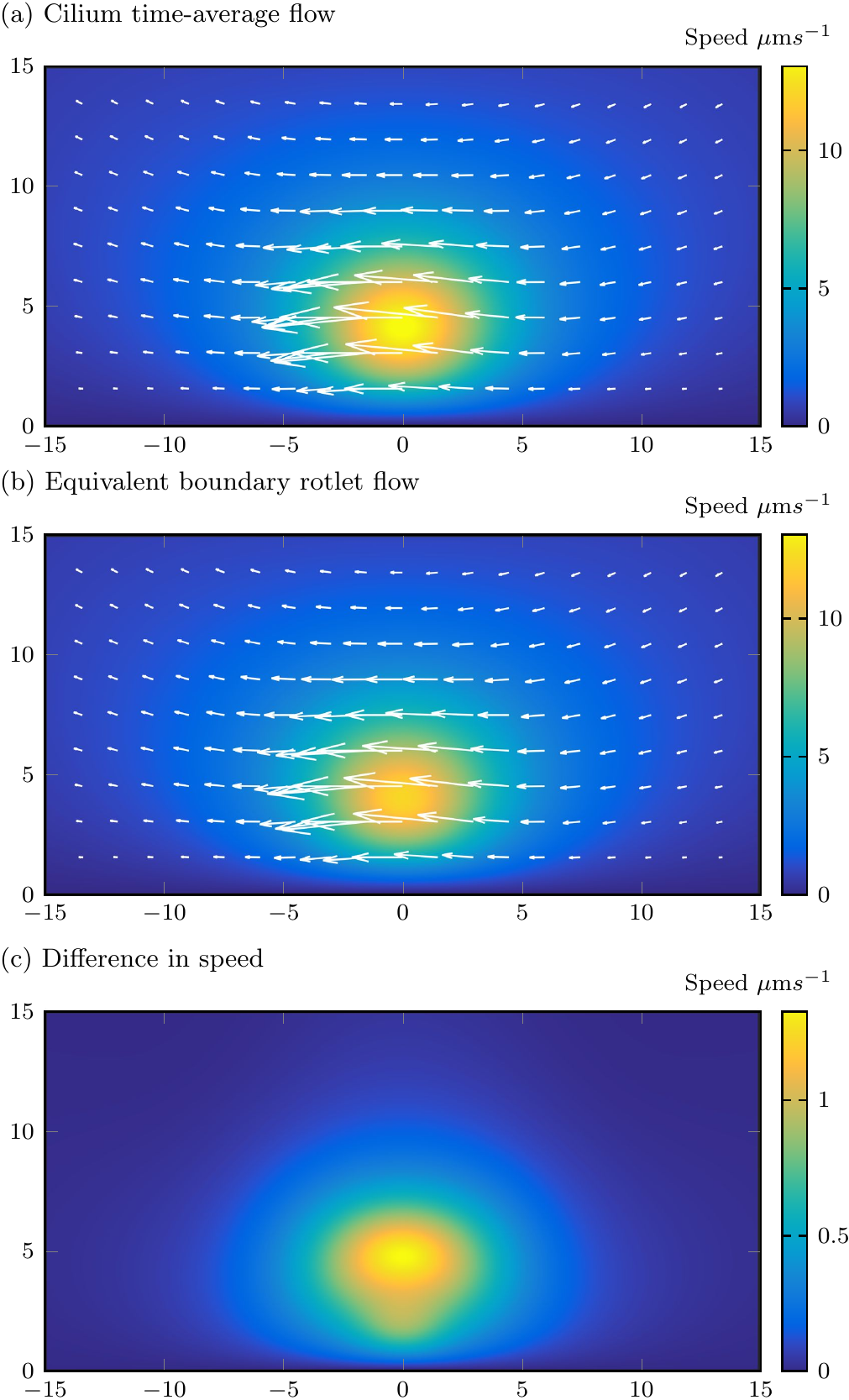}
            \caption{Time averaged flow of the same cilium (a) and equivalent
                rotlet (b) as in figure \ref{fig:z_plane} in the plane $y =
                -2.5\,\mu\mathrm{m}$ half a length away from the cilium. Again (a)
                and (b) are remarkably similar, with (c) showing at worst a $10\%$
                error in the speed calculated by the rotlet model, and very good
                qualitative agreement in flow direction and magnitude.} 
                \label{fig:y_plane}
        \end{center}
    \end{figure}

    Time-averaged flow in the plane $z = 7.5\,\mu\mathrm{m}$ from the average of
    $30$ instants over a single beat as calculated by the resolved cilium model
    is shown in comparison to flow generated by the equivalent rotlet in figure
    \ref{fig:z_plane}. Despite the fact that this plane is only half a cilium
    length away from the cilium tip at its zenith, the equivalent rotlet shows
    at worst a $10\%$ error in the calculated flow, and striking quantitative
    similarity in the flow strength and direction. Similarly, figure
    \ref{fig:y_plane} shows this flow comparison in the plane $y =
    -2.5\,\mu\mathrm{m}$, which is again half a length from the cilium, showing at
    worst a $10\%$ error in the calculated flow. Thus we conclude that even in
    regions relatively near to the cilium (ie greater than half a length), the
    equivalent rotlet provides an accurate description of the flow mechanics.

    \section{Volume flow rate-matching for more complex waveforms}
    \label{sec:flux_match}

    Given a cilium with a more complex centreline parameterisation, for instance
    obtained through high-speed image microscopy, equivalent singularity cilia
    may also be calculated numerically by matching volume flow rates with a time-dependent
    slender body theory cilium model. Beats with helicity induce a net force on
    the fluid in the $\hat{\mathbf{n}}$-direction, in addition to the torque
    component. The volume flow rate per beat induced by a whirling cilium in the
    $\hat{{x}}_1$- and $\hat{{x}}_2$-directions can be matched to
    a plane-boundary rotlet and a blakelet (stokeslet with image singularities)
    using a force-per-unit length extracted from the time-dependent code,
    \begin{subequations}
        \begin{align}
            Q_1 =&\ \left\langle\Int_0^L\frac{f_1(s) \xi_3(s)}{\pi\mu}\,
            \mathrm{d}s \right\rangle = \frac{M\sin{\theta}}{2\pi\mu}, \\
            Q_2 =&\ \left\langle\Int_0^L\frac{f_2(s) \xi_3(s)}{\pi\mu}\,
            \mathrm{d}s \right\rangle = \frac{hF\sin{\theta}}{\pi\mu},
        \end{align}
        \label{eq:cil_centreline_deflected}
    \end{subequations}
    so that
    \begin{equation}
        M = \frac{2}{\sin{\theta}}\left\langle{\Int_0^Lf_1(s) \xi_3(s)\, 
        \mathrm{d}s}\right\rangle, \quad F = \frac{1}{h\sin{\theta}}
        \left\langle{\Int_0^Lf_2(s) \xi_3(s)\, \mathrm{d}s}\right\rangle .
        \label{eq:numerical_flux_matching}
    \end{equation}
    The drawback of this numerical technique is that it requires a slender body
    theory code, and fresh calculation of $M$ and $F$ for different cilium
    lengths and kinematics, which is not necessary with the analytical treatment
    above.  However, an advantage is that this method accounts for non-local
    hydrodynamic interactions in the force calculation, and the numerical method
    may prove increasingly valuable as techniques for three-dimensional waveform
    reconstruction from image microscopy begin to deliver high-resolution
    kinematics \citep{wilson2013high}. 

\end{appendix}

\bibliographystyle{spbasic}
\bibliography{refs}

\end{document}